\documentclass[aps,prb,twocolumn,floatfix,superscriptaddress,showpacs,psfig,reprint, longbilbiography]{revtex4-2}
\usepackage[colorlinks=true,urlcolor=blue,citecolor=blue,linkcolor=blue]{hyperref}
\usepackage{graphicx}
\usepackage{color}
\usepackage{epsfig}
\usepackage{amsmath}
\usepackage{amssymb}
\usepackage{mathrsfs}
\usepackage{bm}
\usepackage{wasysym}
\usepackage{subfigure}
\usepackage{float}
\usepackage{url}
\usepackage{enumerate}
\usepackage{braket}
\usepackage{comment}
\usepackage{multirow}
\usepackage{appendix}
\usepackage[section]{placeins}

\definecolor{violet}{rgb}{0.56,0.0,1.0}

\begin{document}

\hsize\textwidth\columnwidth\hsize\csname@twocolumnfalse\endcsname

\title{Generalized Holstein–Primakoff transformation with optimizable bosonic truncation}

\author{Yang Liu}
\affiliation{School of Physical Science and Technology $\&$ Key Laboratory of Quantum Theory and Applications of MoE, Lanzhou University, Lanzhou 730000, China.}
\affiliation{Lanzhou Center for Theoretical Physics, Key Laboratory of Theoretical Physics of Gansu Province, Lanzhou University, Lanzhou 730000, China.}

\author{Jia-Wei Mei}
\affiliation{Department of Physics, Southern University of Science and Technology, Shenzhen 518055, China}

\author{Rong Yu}
\affiliation{School of Physics, Renmin University of China, Beijing 100872, China} 
\affiliation{Key Laboratory of Quantum State Construction and Manipulation of MoE, Renmin University of China, Beijing 100872, China}

\author{Z. Y. Xie}
\email{qingtaoxie@ruc.edu.cn}
\affiliation{School of Physics, Renmin University of China, Beijing 100872, China}
\affiliation{Key Laboratory of Quantum State Construction and Manipulation of MoE, Renmin University of China, Beijing 100872, China}

\author{Hong-Gang Luo}
\affiliation{School of Physical Science and Technology $\&$ Key Laboratory of Quantum Theory and Applications of MoE, Lanzhou University, Lanzhou 730000, China.}
\affiliation{Lanzhou Center for Theoretical Physics, Key Laboratory of Theoretical Physics of Gansu Province, Lanzhou University, Lanzhou 730000, China.}

\author{Jize Zhao}
\email{zhaojz@lzu.edu.cn}
\affiliation{School of Physical Science and Technology $\&$ Key Laboratory of Quantum Theory and Applications of MoE, Lanzhou University, Lanzhou 730000, China.}
\affiliation{Lanzhou Center for Theoretical Physics, Key Laboratory of Theoretical Physics of Gansu Province, Lanzhou University, Lanzhou 730000, China.}

\begin{abstract}
Spin-wave theory provides a quasiparticle description of excitation spectra in quantum spin systems. In this theory, the spin operators are usually bosonized by 
the Holstein–Primakoff transformation or the Dyson–Maleev transformation. In practical calculations, the bosonized Hamiltonian has to be truncated, and thus 
the resulting excitation spectra depend on the choice of bosonic representation. Here, we introduce a generalized Holstein-Primakoff transformation that 
continuously interpolates between the conventional Holstein–Primakoff and Dyson–Maleev transformations through a single parameter.
The formulation naturally extends to the \(\mathfrak{su}(N)\) algebra and provides a flexible framework for optimizing bosonic truncations beyond harmonic order.
As an application, we investigate the spin-1 bilinear–biquadratic model on the square lattice. By combining the generalized Holstein-Primakoff transformation 
with continuous similarity transformations, we obtain excitation spectra in excellent agreement with tensor-network calculations. Our results
demonstrate that optimizing the bosonic representation substantially reduces truncation errors and provides quantitatively reliable descriptions of both quasiparticle 
dispersions and multi-boson continua.
\end{abstract}

\maketitle
\section{Introduction} 
Spin-wave theory~(SWT)~\cite{Toth_2015} provides one of the most successful microscopic descriptions of collective excitations in dipolar ordered quantum magnets and has now 
become a standard tool for calculating excitation spectra and dynamical response functions. Its central idea is to represent spin operators by bosonic degrees of freedom that 
describe quantum fluctuations around a symmetry-broken reference state. Even just expanding such bosonic operators around the reference state to a linear order, SWT can already
capture the leading quantum fluctuations and yields analytic access to magnon dispersions, Goldstone modes as well as their symmetry properties.
Among various bosonic representations, the Holstein–Primakoff~(HP)~\cite{PhysRev.58.1098} and Dyson–Maleev~(DM)~\cite{PhysRev.102.1217, Maleev1957} transformations are the two 
formulations most widely used in condensed-matter physics. Despite their different mathematical structures, both representations satisfy the same spin algebra and become equivalent 
in the absence of truncation. The generalization of SWT to SU(N), commonly known as generalized spin-wave theory~(GSWT) or flavor-wave theory~\cite{PAPANICOLAOU1988367, PhysRevB.41.9323, PhysRevB.52.3521, PhysRevB.85.140403, 10.1093/ptep/ptu109}, further extends this framework to multipolar ordered phases, spin-orbital systems, and other unconventional quantum magnets.

Despite these successes, quantitative calculations beyond the harmonic approximation remain considerably challenging.
In particular, it has become evident that the accuracy depends crucially on the form of the transformation.
This representation dependence becomes increasingly significant when strong quantum fluctuations generate substantial interactions among bosonic quasiparticles.
In realistic calculations, the bosonized Hamiltonian has to be truncated at a finite order. Consequently, different bosonic representations generally lead to different
effective Hamiltonians after truncation and therefore produce quantitatively different excitation spectra.

In this work, we introduce a generalized Holstein–Primakoff transformation containing a continuous parameter that smoothly interpolates between the conventional HP and DM 
transformations. All members of this family satisfy exactly the same spin algebra and are therefore mathematically equivalent prior to truncation. After truncation, however, 
different choices of the representation parameter lead to different effective Hamiltonians. 
The additional parameter provides a practical degree of freedom for optimizing the truncated theory, reducing representation-induced truncation errors and offering a feasible scheme to improve accuracy.

A second objective of this work is to simplify the implementation of generalized spin-wave theory. Here we reformulate GSWT within an 
automatic-differentiation framework, where the local basis is parameterized by unitary transformations and optimized directly using gradient-based algorithms. This formulation 
is mathematically equivalent to conventional GSWT~\cite{10.1093/ptep/ptu109} while considerably simplifying practical implementations and remaining applicable to arbitrary SU(N) systems.

As an application, we investigate the spin-1 bilinear-biquadratic model on the square lattice. By combining the optimized local basis with the generalized bosonic representation, 
we calculate the excitation spectra and dynamical spin structure factors and benchmark the results against tensor-network calculations.

The remainder of this paper is organized as follows. In Sec.~II we present the theoretical framework, including the automatic-differentiation formulation of generalized spin-wave theory and the generalized Holstein--Primakoff transformation. Numerical results are presented in Sec.~III, followed by conclusions in Sec.~IV.

\section{Method}
\label{METHOD}

\subsection{SU(N) decomposition of a bond Hamiltonian}
Before introducing the framework, we first rewrite the Hamiltonian in a form suitable for a unified SU(N) treatment.
An $N\times{N}$ Hermitian matrix has $N^2$ independent parameters. It can be expressed as a linear combination, with real coefficients, 
of the identity matrix and the generators of SU(N). 
Following the literature~\cite{doi:10.1142/6596}, we introduce three types of traceless Hermitian matrices which are a natural generalization of the Pauli matrices. They are defined as
\begin{align}
\begin{cases}
\left(T_{ab}^{(1)}\right)_{cd} = \dfrac{1}{2} (\delta_{ac}\delta_{bd} + \delta_{bc}\delta_{ad}), \\[1em]
\left(T_{ab}^{(2)}\right)_{cd} = -\dfrac{i}{2} (\delta_{ac}\delta_{bd} - \delta_{bc}\delta_{ad}),
\end{cases}
\end{align}
with $1\le a < b \le N$, and
\begin{equation}
(T_a^{(3)})_{cd} =
\begin{cases}
	\delta_{cd} \{2a(a-1)\}^{-1/2}, & \text{for } c < a, \\[0.5em]
-\delta_{cd} \left( \dfrac{a-1}{2a} \right)^{1/2}, & \text{for } c = a, \\[0.5em]
0, & \text{for } c > a ,
\end{cases}
\label{TA3}
\end{equation}
where \(2 \le a \le N\).

In physics, it is conventional to label these matrices using a single unified index as follows,
\begin{align*}
T^1 &= T_{12}^{(1)}, & T^2 &= T_{12}^{(2)}, & T^3 &= T_{2}^{(3)}, & T^4 &= T_{13}^{(1)}, \\
T^5 &= T_{13}^{(2)}, & T^6 &= T_{23}^{(1)}, & T^7 &= T_{23}^{(2)}, & T^8 &= T_{3}^{(3)}, \\
T^9 &= T_{14}^{(1)}, & T^{10} &= T_{14}^{(2)}, & T^{11} &= T_{24}^{(1)}, & \cdots &
\end{align*}
These \(N^2-1\) traceless Hermitian matrices satisfy the orthogonality relation
\[
\operatorname{Tr}(T^a T^b) = \frac{1}{2}\delta^{ab},
\]
which establishes them as a standard orthogonal basis for the Lie algebra \(\mathfrak{su}(N)\).

As shown in Appendix~\ref{DEC}, omitting the constant and linear terms,
a generic spin Hamiltonian $h_{ij}$ defined on the bond of two sites $i$ and $j$ can be cast into a bilinear form
\begin{equation}
	h_{ij}=\sum_{ab}\mathcal{C}_{ij}^{ab} T^a_iT^b_j.
\end{equation}
This expression makes the underlying SU(N) structure explicit and provides a natural starting point for a unified description of an ordered phase.

\subsection{Automatic differentiation}
\label{RSM}

To incorporate the automatic differentiation technique into the GSWT, we first need to reformulate the Hamiltonian
\begin{eqnarray}
	\mathcal{H} = \sum_{ij}\sum_{ab} \mathcal{C}_{ij}^{ab} T_i^a T_j^b.
	\label{HCIJ}
\end{eqnarray}

By defining $\tilde{T}_i^a = U_i T_i^a U_i^{\dagger}$ with $U_i$ a unitary matrix, we have $T_i^a=U_i^\dagger \tilde{T}_i^a U_i$, which can be expressed as 
a linear combination 
\begin{eqnarray}
	T_i^a=\sum_b \tilde{R}_i^{ab} \tilde{T}_i^b,
	\label{ADJT}
\end{eqnarray}
with $\tilde{R}_i^{ab}=2\mathrm{Tr}\left(U_i^\dagger \tilde{T}_i^a U_i \tilde{T}_i^b \right)$.
Substituting Eq.(\ref{ADJT}) in to (\ref{HCIJ}), we have 
\begin{eqnarray}
	\mathcal{H} = \sum_{ij}\sum_{ab} \tilde{\mathcal{C}}_{ij}^{ab} \tilde{T}_i^a \tilde{T}_j^b.
\end{eqnarray}
For the convenience of calculations, $U_i$s are assumed to be a periodic function on the lattice compatible with the symmetry-broken order.
We express the generators $\tilde{T}_i^a (a=1,2,\cdots, N^2-1)$ in terms of bosonic operators and arrange the Hamiltonian in the normal form. Minimizing the constant 
with respect to $U_i$s, we then obtain optimal $U_i$s. The minimizing procedure is straightforward with automatic differentiation, where the 
classical energy is regarded as a differentiable computational graph. It has been implemented in modern deep-learning libraries 
such as \texttt{PyTorch} or \texttt{LibTorch}~\cite{NEURIPS2019_bdbca288}, allowing efficient optimization even when a large number of variational parameters are involved.

The resulting formulation is mathematically equivalent to conventional GSWT but substantially simplifies practical implementations. Moreover, the same optimization framework 
applies without modification to arbitrary values of N .

In practice, it is more convenient to formulate the $U_i$s as $U=\exp(-iK)$ with~\cite{doi:10.1142/6596} 
\begin{eqnarray}
	K=\sum_{a<b}\{w^{(1)}_{ab}T_{ab}^{(1)}+w^{(2)}_{ab}T^{(2)}_{ab}\}+\sum_{a=2}^N w^{(3)}_{a}T_a^{(3)},
\end{eqnarray}
here $w^{(1)}_{ab}$, $w^{(2)}_{ab}$ and $w^{(3)}_{a}$ are variational parameters to be optimized. This parametrization automatically preserves unitarity during optimization. 

\subsection{Generalized Holstein-Primakoff transformation}
In SWT, the spin-$S$ operators are commonly mapped into bosonic operators by the HP transformation~\cite{PhysRev.58.1098}
\begin{align}
S^z&=S-a^\dagger{a}, \nonumber \\
S^+&=\left(2S-a^\dagger{a}\right)^{1/2}a, \nonumber \\
S^-&=a^\dagger\left(2S-a^\dagger{a}\right)^{1/2},
\end{align}
or alternatively, by the non-Hermitian DM transformation~\cite{PhysRev.102.1217, Maleev1957}
\begin{align}
S^z&=S-a^\dagger{a}, \nonumber \\
S^+&=\left(2S-a^\dagger{a}\right)a, \nonumber \\
S^-&=a^\dagger,
\end{align}
where $a$ and $a^\dagger$ are bosonic operators obeying the relation $[a,a^\dagger]=1$ and the constraint $a^\dagger a \le 2S$.
Although these two transformations have rather different algebraic forms, they satisfy exactly the same spin commutation relations and are therefore completely
equivalent before any approximation is introduced.

Motivated by this observation, we introduce the generalized HP transformation
\begin{align}
S^z&=S-a^\dagger{a}, \nonumber \\
S^+&=\left(2S-a^\dagger{a}\right)^{x}a, \nonumber \\
S^-&=a^\dagger\left(2S-a^\dagger{a}\right)^{1-x},
\label{GHPT}
\end{align}
where $0\le{x}\le{1}$. 

One can verify that Eq.~(\ref{GHPT}) satisfies the exact spin algebra for arbitrary $x$. 
The conventional HP transformation is recovered at $x=1/2$, while $x=0$ and $x=1$ correspond to the two equivalent forms of the DM transformation.
To our knowledge, an explicit interpolating form of this kind has not been employed in practical spin-wave calculations; 
the underlying factorization freedom was noted in Ref.~\cite{tkeshelashvili2016bosonrepresentationspinoperators}, but no explicit parameterization was given there.
In the following sections, we will extend the generalized Holstein-Primakoff transformation~(\ref{GHPT}) to the SU(N) case when discussing specific models.

\subsection{Optimization of bosonic truncation}
In general, the bosonic Hamiltonian can be written as 
\begin{equation}
\mathcal{H} = \mathcal{H}_0 + \mathcal{H}_2 + \mathcal{H}_3 + \mathcal{H}_4 + \mathcal{H}_5 +\cdots,
\end{equation}
where $\mathcal{H}_n$ contains terms involving $n$ bosonic operators.

One may notice that for the square-lattice spin-1/2 Heisenberg model, the Dyson–Maleev transformation ($x = 0$ or $1$ in Eq.~(\ref{GHPT})) yields the most accurate 
results~\cite{PhysRevLett.115.207202,10.21468/SciPostPhys.4.1.001}. This is because the resulting Hamiltonian contains at most quartic terms. For a general model, 
no such simple choice exists. In practical calculations, however, the bosonized Hamiltonian must be truncated at finite order, typically to $\mathcal{H}_4$ in interacting SWT and GSWT.
Consequently, different values of $x$ generally produce different truncated Hamiltonians. The parameter $x$ therefore provides an additional degree of freedom for improving the quantitative accuracy of interacting spin-wave calculations. In this sense, $x$ plays a role analogous to a variational parameter whose optimal value depends on 
the truncation scheme and the physical system.

\section{Results}

To demonstrate the capability of the proposed framework, we apply it to the spin-1 bilinear-biquadratic~(BBH) model on the square lattice. This model
provides an ideal benchmark because it hosts both conventional magnetic phases and unconventional quadrupolar order~\cite{PAPANICOLAOU1988367, PhysRevB.65.052403, PhysRevB.87.224431, SciPostPhys.3.4.030}. Throughout this section, we focus on the ferroquadrupolar~(FQ) phase and systematically compare the proposed automatic-differentiation 
generalized spin-wave theory~(aGSWT) with conventional GSWT~\cite{10.1093/ptep/ptu109} and TN calculations~\cite{PhysRevB.101.195109, 10.21468/SciPostPhys.12.1.006, PhysRevX.9.031041}.

The discussion proceeds in three steps. We first validate the proposed aGSWT formulation at the harmonic level by comparing it with the conventional GSWT.
We then identify the limitations of the harmonic approximation through comparisons with TN results. Finally, we demonstrate that optimizing the generalized Holstein-Primakoff 
representation substantially improves the description of interacting excitations.

\subsection{Spin-1 bilinear-biquadratic model}
The spin-1 BBH model on the square lattice is given by the Hamiltonian
\begin{equation}
	\mathcal{H}=\sum_{\langle{i,j}\rangle}h_{ij}, \label{BBH}
\end{equation}
where 
\begin{equation}
	h_{ij}=J_1\bm{S}_i \cdot \bm{S}_j+J_2(\bm{S}_i \cdot \bm{S}_j)^2.
\end{equation}
The coupling constants $J_1$ and $J_2$ can be parameterized as $J_1=\cos{\theta}$ and $J_2=\sin{\theta}$.  
The competition between the bilinear and biquadratic interactions gives rise to a rich phase diagram~\cite{PAPANICOLAOU1988367, PhysRevB.65.052403, PhysRevB.87.224431, SciPostPhys.3.4.030} containing ferromagnetic~(FM), FQ, and antiferromagnetic~(AFM) phases. In this work we concentrate on the FQ phase.

$\bm{S}_i$ is the spin-1 operator at site $i$, with its three components written as 
\begin{equation}
S^x_i=\frac{1}{\sqrt{2}}\begin{pmatrix} 0 & 0 & 1 \\ 0 & 0 & 1 \\ 1 & 1 & 0 \end{pmatrix},
\end{equation}
\begin{equation}
S^y_i=\frac{i}{\sqrt{2}}\begin{pmatrix} 0 & 0 & -1 \\ 0 & 0 & 1 \\ 1 & -1 & 0 \end{pmatrix}
\end{equation}
and
\begin{equation}
S^z_i=\begin{pmatrix} 1 & 0 & 0 \\ 0 & -1 & 0 \\ 0 & 0 & 0 \end{pmatrix}.
\end{equation}
Here, the local basis states of the three matrices are ordered in terms of the eigenvalues of $S^z_i$, specified by $|1\rangle$, $|-1\rangle$ and $|0\rangle$. 

To explore the excitations, it is convenient to rewrite the Hamiltonian $h_{ij}$ in terms of the generators of SU(3), i.e., $T_i^a$ and $T_j^a$. It reads 
%\begin{widetext}
%\begin{equation}
%\begin{split}
%        h_{ij} & = 2(2J_1-J_2) T_i^3 T_j^3 + 2J_2 T_i^8 T_j^8 + 2J_2 T_i^1 T_j^1 + 2J_2 T_i^2 T_j^2 + 2J_1 T_i^4 T_j^4 + 2J_1 T_i^5 T_j^5 + 2J_1 T_i^6 T_j^6 + 2J_1 T_i^7 T_j^7 \\
%        & \quad + 2(J_1-J_2)\left(T_i^4 T_j^6 + T_i^6 T_j^4 - T_i^7 T_j^5 - T_i^5 T_j^7 \right)\\ 
%        & \quad + \frac{4}{3}J_2.
%        \label{HIJ}
%\end{split}
%\end{equation}
%\end{widetext}
\begin{equation}
 \begin{aligned}
          h_{ij} & = 2(2J_1-J_2) T_i^3 T_j^3 + 2J_2 T_i^8 T_j^8 + 2J_2 T_i^1 T_j^1 + 2J_2 T_i^2 T_j^2 \\
	         & \quad + 2J_1 T_i^4 T_j^4 + 2J_1 T_i^5 T_j^5 + 2J_1 T_i^6 T_j^6 + 2J_1 T_i^7 T_j^7 \\
        & \quad + 2(J_1-J_2)\left(T_i^4 T_j^6 + T_i^6 T_j^4 - T_i^7 T_j^5 - T_i^5 T_j^7 \right)\\
        & \quad + \frac{4}{3}J_2.
        \label{HIJ}
 \end{aligned}
\end{equation}

Defining a unitary transformation~(see also Appendix~\ref{appRDM}) $\tilde{T}_i^a = U_i T_i^a U_i^{\dagger}$, we can rewrite the Hamiltonian in terms of $\tilde{T}_i^a$ following 
the standard procedure given in Sec.~\ref{RSM}.
The next step is to bosonize the generators $\tilde{T}_i^a$ so that the fluctuation is governed by the non-Cartan generators $\tilde{T}_i^a$ with $a = 1, 2, 4, 5$ and $7$.
The details are given in Appendix~\ref{bosonRep}. After expressing the Hamiltonian~(\ref{BBH}) by the bosonic operators, minimizing the constant term using the automatic-differentiation procedure with respect to the $U_i$s leads to the final optimal Hamiltonian.

\subsection{Excitation spectra in the FQ phase}
\begin{figure}[htb]
        \centering
        \includegraphics[width=0.95\columnwidth, clip]{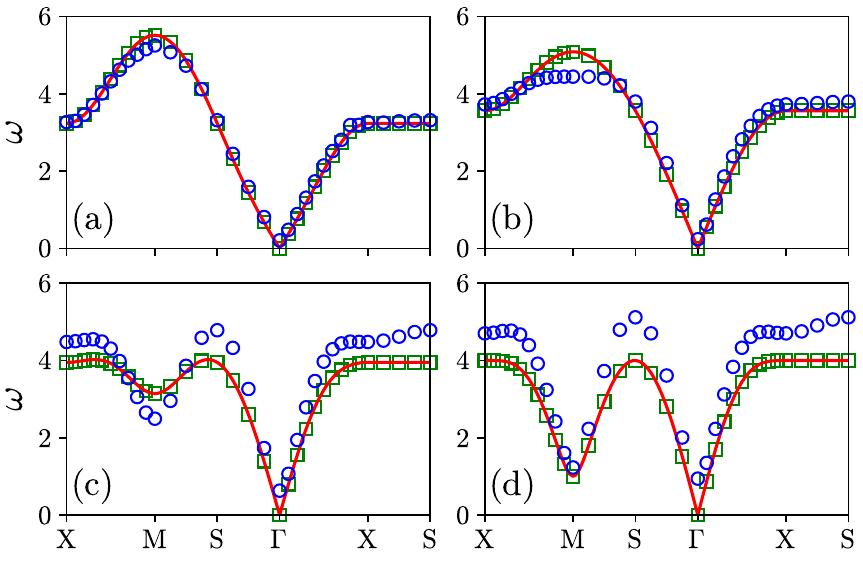}
        \caption{The dispersion relation of the quadrupolar mode along the path $\mathrm{X}-\mathrm{M}-\mathrm{S}-\mathrm{\Gamma}-\mathrm{X}-\mathrm{S}$.
        The squares ({$\square$}) and circles ($\circ$) are obtained by linear aGSWT and TN methods, respectively. The solid lines are plotted
        according to the appendix of Ref.~\cite{PhysRevB.87.224431}. The panels from (a) to (d) correspond to $\theta = 1.3\pi, 1.35\pi, 1.45\pi$ and $1.495\pi$.}
        \label{FIG1}
\end{figure}
The FQ phase occupies the regime $5\pi/4<\theta<3\pi/2$ in the phase diagram. At $\theta=5\pi/4$, it undergoes a phase transition into the FM phase,
while at $\theta=3\pi/2$ it undergoes a transition into the AFM phase~\cite{SciPostPhys.3.4.030}.
In contrast to conventional magnetically ordered phases, the FQ phase is characterized by a uniform alignment of quadrupolar directors while maintaining a vanishing dipolar moment~\cite{PAPANICOLAOU1988367, SciPostPhys.3.4.030}, i.e., $\langle \mathbf{S}_i \rangle = 0$. The corresponding order parameter is a traceless rank-2 tensor~\cite{PhysRevB.82.174440}, reflecting the nematic nature of the state. 
Consequently, the low-energy excitations are predominantly quadrupolar. Despite the absence of magnetic order, the FQ phase spontaneously breaks the global SO(3) symmetry, leading to the emergence of gapless collective excitations.

\subsubsection{Validation of harmonic aGSWT and its limitations}

We first examine the harmonic approximation in order to validate the proposed automatic-differentiation formulation. Figure~\ref{FIG1} compares the excitation spectra obtained by 
linear aGSWT with those from conventional GSWT (see, e.g., the appendix of Ref.~\cite{PhysRevB.87.224431}) and TN calculations.
The dispersion obtained by linear aGSWT is identical to that of the conventional GSWT over the entire Brillouin zone, confirming that the present formulation introduces no additional 
approximation. The automatic-differentiation optimization therefore reproduces exactly the harmonic theory.
This demonstrates that the proposed formulation is mathematically equivalent to the conventional GSWT but considerably simpler to implement.

The comparison with the TN calculations further reveals the range of validity of the harmonic approximation. Near the FM-FQ transition [$\theta=1.3\pi$, Fig.~1(a)],
the harmonic theory already provides an excellent description of the low-energy dispersion. However, systematic deviations emerge as the
system approaches the FQ-AFM transition. For $\theta=1.45\pi$ and $1.495\pi$, in particular, the TN results exhibit a weak softening around the $X$ point, resulting in a roton-like 
minimum~\cite{PhysRevLett.115.207202, PhysRevX.7.041072, PhysRevB.111.245117} that is completely absent within the harmonic approximation. 
The discrepancy indicates that interactions among bosonic quasiparticles play a growing role in shaping the excitation spectrum.
\begin{figure*}[hbtp!]
        \includegraphics[width=0.99\textwidth, clip]{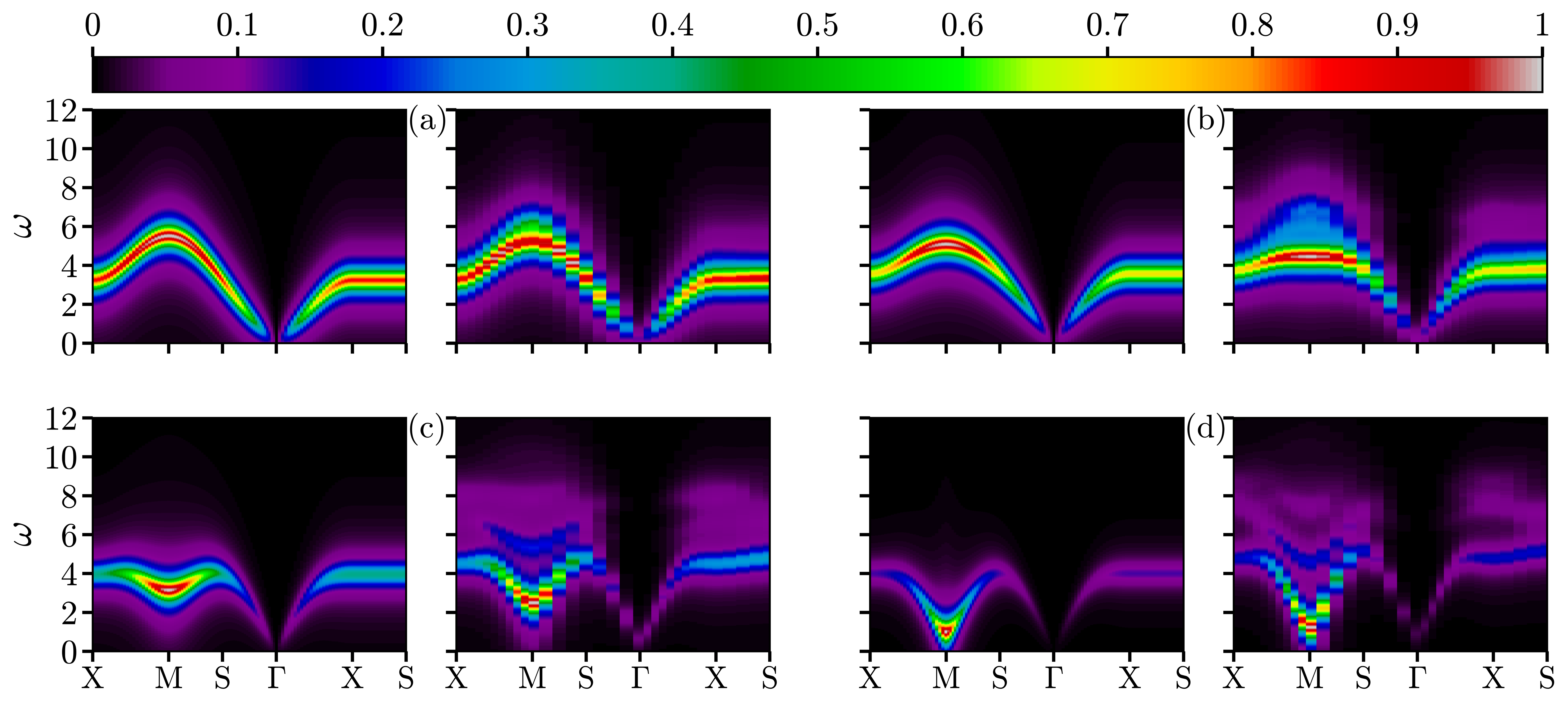}
	\caption{$A_{\bm{k} \omega} (S^{-}, S^{+})$ are shown for $\theta=1.3\pi$(a), $1.35\pi$(b), $1.45\pi$(c) and $1.495\pi$(d). For each $\theta$, the left and right subfigures 
	are obtained by the linear aGSWT and TN methods, respectively.}
        \label{FIG2}
\end{figure*}

To establish a direct connection with experimental observables, we evaluate the dynamical structure factor (or spectral function), which encodes both the excitation energies and 
their spectral weights measured directly in inelastic neutron-scattering experiments. It is defined as
\begin{equation}
\begin{split}
A_{\bm{k} \omega}(\hat{X}, \hat{Y})=\sum_{n>0}\frac{\langle 0|\hat{X}_{-\bm{k}}|R_n\rangle\langle L_n| \hat{Y}_{\bm{k}}|0\rangle}{\langle L_n| R_n\rangle}\delta(\omega-\omega_n).
\end{split} 
\end{equation}
Here, $\langle L_n|$ and $|R_n\rangle$ are the $n$-th left and right eigenvectors, $\omega_n$ is the $n$-th excitation energy, and $\langle L_n| = |R_n\rangle^{\dag}$ if the Hamiltonian is 
Hermitian. The operator $\hat{X}_{\bm{k}}$ in $\bm{k}$-space is the Fourier transform of $\hat{X}_i$ in real space, given by $\hat{X}_{\bm{k}}=\frac{1}{L^2}\sum_{i}e^{i\bm{k} \cdot \bm{r}_i}\hat{X}_i$ with $L$ the linear size of the square lattice. In practical calculations, the $\delta$ function is replaced by a Lorentzian distribution with a broadening $0.5$ to 
mimic experimental resolution. In this study, we focus on the dynamic spin structure factors $A_{\bm{k} \omega}(S^{-}, S^{+})$, 
which are accessible via inelastic neutron scattering.

Figure~\ref{FIG2} compares $A_{\bm{k} \omega} (S^{-}, S^{+})$ obtained from the linear aGSWT and the TN calculations for four representative $\theta$. Both methods 
exhibit the same overall evolution as $\theta$ increases, confirming that the harmonic theory correctly captures the qualitative nature of the excitations.

The quadrupolar nature of the Goldstone mode has a direct spectral consequence: the spectral weight 
of $A_{\bm{k} \omega} (S^{-}, S^{+})$ at the $\Gamma$ point is vanishingly small. 
This qualitative distinction from conventional magnon excitations plays a central role in both the theoretical characterization and 
experimental identification of the FQ phase. 
Meanwhile, the spectral weight around the $M$ point grows steadily as $\theta$ approaches the FQ-AFM transition, accompanied by a continuous softening of the excitation gap.
These collectively suggest that the system progressively approaches the antiferromagnetic limit as $\theta$ increases.

Although linear aGSWT successfully reproduces the overall dispersion and the low-energy spectral distribution, significant quantitative differences emerge at larger values of $\theta$.
Most notably, the TN spectra exhibit a broad high-energy continuum with no counterpart in the harmonic approximation. 
The continuum gains spectral weight near the phase boundary, at the expense of the coherent one-particle branch and through redistribution into multi-particle excitations.
Such redistribution is a direct manifestation of strong interactions between bosonic quasiparticles and therefore cannot be described within a purely quadratic bosonic Hamiltonian.

\subsubsection{Optimization of the bosonic representation}
%\FloatBarrier
\begin{figure*}[hbt]
        \includegraphics[width=0.99\textwidth, clip]{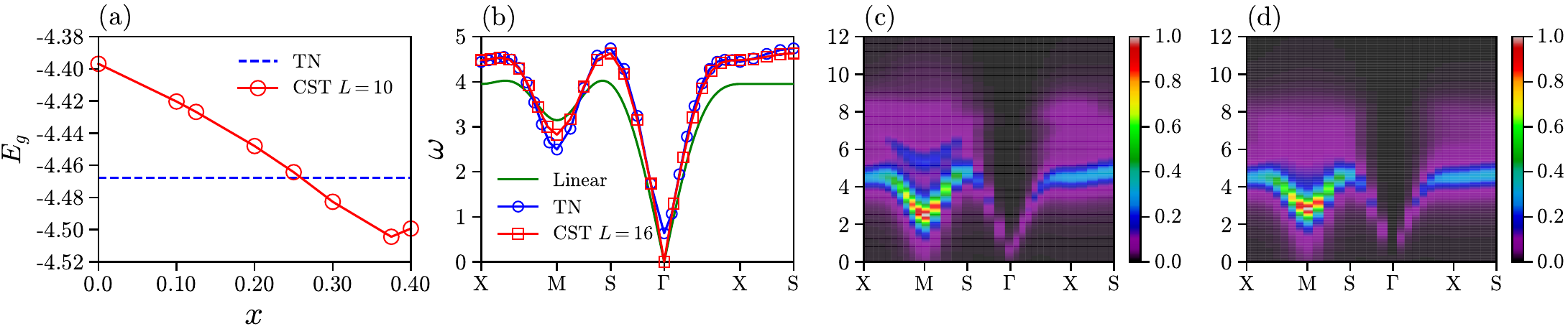}
        \caption{(a) The ground-state energy $E_g$ at $\theta=1.45\pi$ obtained by CST is shown as a function of $x$. The TN result extrapolated to the thermodynamic limit 
	is plotted as a benchmark.
		 (b) The dispersion curves of quadrupolar mode obtained by linear aGSWT, TN, and CST methods are plotted along the path $\mathrm{X}$-$\mathrm{M}$-$\mathrm{S}$-$\mathrm{\Gamma}$-$\mathrm{X}$-$\mathrm{S}$.
		 (c),(d) $A_{\bm{k} \omega} (S^{-}, S^{+})$ are shown, which are obtained by the TN and CST~($L=16$) methods, respectively.}
        \label{rfig4}
\end{figure*}
The comparisons above make clear that the harmonic approximation is no longer sufficient near the FQ–AFM phase boundary. A quantitative description therefore requires going beyond the harmonic 
approximation by incorporating boson–boson interactions. In the present work, this is achieved by combining the generalized HP 
transformation with the continuous similarity transformation~(CST)~\cite{PhysRevLett.115.207202,10.21468/SciPostPhys.4.1.001}.

In practical CST calculations, the bosonic Hamiltonian must be truncated at finite order, typically retaining terms up to quartic order.  
An optimal choice of $x$ in the generalized HP transformation~(\ref{GHPT}) should minimize the contribution from the discarded terms. However, determining an optimal $x$
rigorously is technically challenging. In particular, $x$ is not a variational parameter. Instead, we adopt a feasible strategy by benchmarking against quantities 
that are either known or easily accessible. The most straightforward choice is the ground-state energy. Figure~\ref{rfig4}(a) shows the CST ground-state energy as a function of $x$ at $\theta=1.45\pi$, 
together with the tensor-network result. Around $x\simeq0.25$, the CST calculation agrees remarkably well with the TN benchmark. Unless otherwise stated, this optimized value 
is adopted for all subsequent calculations at the same parameter point.

The optimized representation leads to a substantial improvement in the excitation spectrum. As shown in Fig.~\ref{rfig4}(b), the dispersion obtained by CST closely follows the TN result throughout the Brillouin zone. In particular, the interaction-induced roton-like minimum around the X point--missed entirely by the harmonic theory--is successfully reproduced after optimizing the representation parameter.
The agreement indicates that the optimized generalized Holstein--Primakoff transformation effectively suppresses the truncation error introduced by the finite-order bosonic expansion.

A more stringent benchmark is provided by the dynamical spin structure factor. Figures~\ref{rfig4}(c) and \ref{rfig4}(d) compare the spectra obtained by TN and the CST calculation. Besides reproducing the coherent one-particle branch, the optimized theory also captures the broad high-energy continuum arising from multi-boson excitations~\cite{PhysRevB.79.144416, RevModPhys.85.219}. Both the spectral distribution and the overall transfer of spectral weight agree well with the TN results, demonstrating that the optimization improves not only the quasiparticle dispersion but also the underlying many-body dynamics.

Some quantitative differences nevertheless remain. In particular, the TN calculation exhibits a weak secondary feature around the $M$ point that is absent in the present CST spectra. 
This discrepancy is most likely associated with the present implementation of CST, where only the two-particle Hilbert space is retained. Including higher-particle sectors is expected to further improve the description of the high-energy continuum without affecting the optimization strategy proposed here.

The simultaneous improvement of the ground-state energy, excitation spectrum, and dynamical structure factor demonstrates that optimizing the bosonic representation systematically reduces the truncation error of interacting spin-wave calculations. Rather than representing a simple interpolation between the conventional HP and DM transformations, the generalized transformation provides a continuous family of equivalent exact representations whose optimal member depends on the underlying Hamiltonian and the adopted approximation scheme. 
Moreover, our scheme is compatible with the recently proposed resummation of the HP expansion~\cite{MichaelPRR2020} and projected HP representation~\cite{Liuke2026}, 
and combining these may further reduce truncation errors.
This viewpoint considerably enhances the flexibility of interacting spin-wave theory while remaining fully compatible with the exact spin algebra.

\section{Conclusion}

In this work, we have extended GSWT in two complementary directions.
First, we reformulated GSWT within an automatic-differentiation framework, which allows the optimization procedure to be readily incorporated into a modern 
deep-learning library. Second, we introduced a generalized HP transformation that continuously interpolates between the conventional HP 
and DM representations, enabling the bosonic representation to be treated as an adjustable object after finite-order truncation.

The proposed approach was applied to the spin-1 bilinear-biquadratic model on the square lattice. At the harmonic level, the automatic-differentiation formulation reproduces 
exactly the results of conventional GSWT, confirming the validity of the numerical optimization procedure.
Beyond the harmonic approximation, combining the generalized HP transformation with CST significantly improves the quantitative 
agreement with tensor-network benchmarks. In particular, the optimized representation correctly captures the roton-like feature in the excitation spectrum and the multi-boson continuum 
in the dynamical spin structure factor.

A central outcome of this work is that the choice of bosonic representation becomes physically relevant once the Hamiltonian is truncated at finite order.
Although different representations are exactly equivalent in the full theory, they generally lead to different effective interacting Hamiltonians after truncation.
This observation suggests that the bosonic representation can be regarded as another degree of freedom that can be optimized to reduce truncation errors.
We expect that this optimizable bosonization scheme will find applications in a wide range of strongly correlated quantum systems.
In addition, since one can construct many bosonic representations satisfying the \(\mathfrak{su}(N)\) algebra, an optimizable bosonic representation may be model-dependent under truncation. 
How to choose a feasible representation deserves further exploration.

\section{Acknowledgements.}
We are indebted to Hong-Hao Tu for the discussions that inspired this work. We are also grateful to Gang Chen for his helpful discussions.
This work is supported by the National key R$\&$D Program of China (Grants Nos. 2022YFA1402704, 2024YFA1408604, 2023YFA1406500, 2021YFA1400400), 
by the National Natural Science Foundation of China (Grants Nos. 12274187, 12274458, 12334008, 12474143, 12247101), by the Shenzhen Fundamental Research Program (Grant Nos. JCYJ20220818100405013 and JCYJ20230807093204010) and by the Guangdong Provincial Quantum Science Strategic Initiative (Grant No. GDZX2401010).

\appendix
\section{Hamiltonian in terms of SU(N) generators}
\label{DEC}

Any operator $O$ can be decomposed into two operators as
\begin{equation}
O = \frac{1}{2}(X+iY),
\label{ODEC}
\end{equation}
where
\begin{equation}
X = O + O^\dagger, \qquad \nonumber
Y = \frac{1}{i}(O - O^\dagger).  \nonumber
\end{equation}
By construction, both $X$ and $Y$ are Hermitian operators satisfying
\begin{equation}
X^\dagger = X, \qquad Y^\dagger = Y.  \nonumber
\end{equation}

On the other hand, a generic spin Hamiltonian defined on the bond of two sites $i$ and $j$ can be written in the form
\begin{equation}
h_{ij} = O_i O_j^\dagger + O_i^\dagger O_j,
\label{HBIJ}
\end{equation}
where $O_i$ denotes a local operator acting on site $i$. Substituting the decomposition~(\ref{ODEC})
\begin{equation}
O_i = \frac{1}{2}(X_i + iY_i),
\qquad
O_i^\dagger = \frac{1}{2}(X_i - iY_i)
\end{equation}
into the bond Hamiltonian~(\ref{HBIJ}), one obtains
\begin{equation}
h_{ij} = \frac{1}{2}\left( X_i X_j + Y_i Y_j \right),
\end{equation}
which explicitly expresses $h_{ij}$ as a summation of products of Hermitian operators. Moreover, $X_i$ and $Y_i$ are linear combinations of
generators of SU(N) plus the identity matrix. Thus, omitting the constant and linear terms, any $h_{ij}$ can be cast into a bilinear form
\begin{equation}
        h_{ij}=\sum_{ab}\mathcal{C}_{ij}^{ab} T^a_iT^b_j.
\end{equation}

\section{SU(3) generators and their bosonization}
\label{bosonRep}

To be specific, the generators of the \(\mathfrak{su}(3)\) Lie algebra in the fundamental representation read
\begin{equation}
\begin{aligned}
T^1 &= \frac12
\begin{pmatrix}
0 & 1 & 0 \\
1 & 0 & 0 \\
0 & 0 & 0
\end{pmatrix},
&
T^2 &= \frac12
\begin{pmatrix}
0 & -i & 0 \\
i & 0 & 0 \\
0 & 0 & 0
\end{pmatrix}, \\[6pt]
T^3 &= \frac12
\begin{pmatrix}
1 & 0 & 0 \\
0 & -1 & 0 \\
0 & 0 & 0
\end{pmatrix},
&
T^4 &= \frac12
\begin{pmatrix}
0 & 0 & 1 \\
0 & 0 & 0 \\
1 & 0 & 0
\end{pmatrix}, \\[6pt]
T^5 &= \frac12
\begin{pmatrix}
0 & 0 & -i \\
0 & 0 & 0 \\
i & 0 & 0
\end{pmatrix},
&
T^6 &= \frac12
\begin{pmatrix}
0 & 0 & 0 \\
0 & 0 & 1 \\
0 & 1 & 0
\end{pmatrix}, \\[6pt]
T^7 &= \frac12
\begin{pmatrix}
0 & 0 & 0 \\
0 & 0 & -i \\
0 & i & 0
\end{pmatrix},
&
T^8 &= \frac{1}{2\sqrt3}
\begin{pmatrix}
1 & 0 & 0 \\
0 & 1 & 0 \\
0 & 0 & -2
\end{pmatrix}.
\end{aligned}
\end{equation}
where they satisfy
\begin{equation}
T^a = \frac{\lambda^a}{2}, \qquad a=1,\dots, 8,  \nonumber
\end{equation}
with $\lambda^a$ the well-known Gell-Mann matrices.

The common eigenvectors of two Cartan generators $T_i^3$ and $T_i^8$ are $|w_1\rangle=(1,0,0)^T$, $|w_2\rangle=(0,1,0)^T$ and $|w_3\rangle=(0,0,1)^T$.
The corresponding three weights are $w_1=(\frac{1}{2}, \frac{1}{2\sqrt{3}})$, $w_2=(\frac{-1}{2}, \frac{1}{2\sqrt{3}})$, and $w_3=(0, \frac{-1}{\sqrt{3}})$
with $w_1$ being the highest weight.

To describe the fluctuations clearly, it is more convenient to introduce the Cartan-Weyl basis,
which includes $T_{i}^3$, $T_{i}^8$, $E_{i}^{\pm\alpha_1}$, $E_{i}^{\pm\alpha_2}$, and $E_{i}^{\pm(\alpha_1+\alpha_2)}$
with $E_{i}^{\pm\alpha_1}=T_{i}^{4}\pm iT_{i}^{5}$, $E_{i}^{\pm\alpha_2}=T_{i}^{6}\mp iT_{i}^{7}$ and $E_{i}^{\pm(\alpha_1+\alpha_2)}=T_{i}^{1}\pm iT_{i}^{2}$.
The action of $E_{i}^{\alpha}$($\alpha=\pm\alpha_1,\pm\alpha_2,\pm(\alpha_1+\alpha_2)$ with the simple roots of $\mathfrak{su}(3)$ $\alpha_1=(\frac{1}{2}, \frac{\sqrt{3}}{2})$ and $\alpha_2=(\frac{1}{2}, \frac{-\sqrt{3}}{2}$) on one of these eigenvectors $|w_m\rangle(m=1,2,3)$ is
\begin{equation}
E_{i}^{\alpha}|w_m\rangle=|w_m+\alpha\rangle, 
\end{equation}
transforming it into an eigenvector with weight $w_m+\alpha$. If $w_m+\alpha$ is not a valid weight, then $E_{i}^{\alpha}|w_m\rangle=0$.
They satisfy $\alpha_1=w_1-w_3$, $\alpha_2=w_3-w_2$ and $\alpha_1+\alpha_2=w_1-w_2$.

After the unitary transformation $\tilde{T}_i^a=U_iT_i^a U_i^\dagger$, the Cartan-Weyl basis can be bosonized by extending the generalized Holstein-Primakoff transformation~(\ref{GHPT}) to SU(3) case, which reads
\begin{equation}
\begin{split}
	&\tilde{T}_{i}^{3}=\frac{1}{2}\left(1-2b_i^{\dag}b_i-c_i^{\dag}c_i\right), \\
	&\tilde{T}_{i}^{8}=\frac{1}{2\sqrt{3}}\left(1-3c_i^{\dag}c_i\right),\\
	&\tilde{E}_{i}^{\alpha_1+\alpha_2}=\left(1-b_i^{\dag}b_i-c_i^{\dag}c_i\right)^{x}b_i,\\
	&\tilde{E}_{i}^{-(\alpha_1+\alpha_2)}=b_i^{\dag}\left(1-b_i^{\dag}b_i-c_i^{\dag}c_i\right)^{1-x},\\
	&\tilde{E}_{i}^{\alpha_1}=\left(1-b_i^{\dag}b_i-c_i^{\dag}c_i\right)^{x}c_i,\\
	&\tilde{E}_{i}^{-\alpha_1}=c_i^{\dag}\left(1-b_i^{\dag}b_i-c_i^{\dag}c_i\right)^{1-x},\\
	&\tilde{E}_{i}^{\alpha_2}=c_i^{\dag}b_i, \\
	&\tilde{E}_{i}^{-\alpha_2}=b_i^{\dag}c_i.
\end{split}\label{ghp}
\end{equation}
Here, $b_i$, $b_i^{\dag}$, $c_i$, and $c_i^{\dag}$ are bosonic operators under the constraint $b_i^\dagger b_i+c_i^\dagger c_i\le 1$,
representing quantum fluctuations around the reference state. Notably, when $x=1/2$, they become the SU(3) HP transformation and when $x=0$ or $x=1$,
they turn into the SU(3) Dyson-Maleev transformation. Their extension to SU(N) is straightforward so we will not elaborate further here.

\section{Reduced density matrix}
\label{appRDM}
In some cases, these $U_i$s may be calculated from the ground state $|\psi_g\rangle$, or equivalently, the reduced density matrix $\rho_i$ for site $i$. which are obtained by the TN method.  
In terms of $\rho_i$, the expectation values $\langle T_i^a\rangle = \langle\psi_g|T_i^a|\psi_g\rangle/\langle\psi_g|\psi_g\rangle$
can be written as $\langle T_i^a\rangle=\mathrm{Tr}\left(\rho_i T_i^a\right)$.
Since $\rho_i$ is a Hermitian matrix, it can be unitarily diagonalized, i.e.,
$\rho_i=U_i\Gamma_i U_i^\dagger$. Here the diagonal elements of $\Gamma_i$ are in descending order. By defining $\tilde{T}_i^a=U_i T_i^a U_i^\dagger$, we have
\begin{equation}
        \langle\tilde{T}_{i}^{a}\rangle=\mathrm{Tr}(\rho_i\tilde{T}_{i}^{a})=\mathrm{Tr}(\Gamma_i T_{i}^{a})=\sum_m \Gamma_i^{mm} T^a_{i,{mm}},
\label{LAME}
\end{equation}
which indicates that only $\langle \tilde{T}^{(3)}_{a,i}\rangle$$(2\le{a}\le{N})$ may acquire a finite value and all others are zero. The common eigenvectors of the Cartan generators 
$\tilde{T}^{(3)}_{a,i}$ are
\begin{align}
|w_1\rangle = U_i
\begin{pmatrix}
1 \\ 0\\ \vdots \\ 0
\end{pmatrix},
|w_2\rangle = U_i
\begin{pmatrix}
0 \\ 1\\ \vdots \\ 0
\end{pmatrix},
\cdots,
|w_N\rangle = U_i
\begin{pmatrix}
0 \\ 0\\ \vdots \\ 1
\end{pmatrix}. \nonumber
\end{align}
The corresponding eigenvalues written in vectors are
\begin{align}
w_k =
\begin{cases}
\left(\dfrac{1}{\sqrt{4}}, \dfrac{1}{\sqrt{12}}, \cdots, \dfrac{1}{\sqrt{2N(N-1)}}\right), & k=1 \\[6pt]
\left(-\dfrac{1}{\sqrt{4}}, \dfrac{1}{\sqrt{12}}, \cdots, \dfrac{1}{\sqrt{2N(N-1)}}\right), & k=2 \\[6pt]
\;\; \vdots \\[6pt]
\left(0, 0, \cdots, -\dfrac{N-1}{\sqrt{2N(N-1)}}\right), & k=N
\end{cases} \nonumber
\end{align}
They are known as the weights of the \(\mathfrak{su}(N)\) Lie algebra, with $w_1$ being the highest weight. 
If $(\langle\tilde{T}_{2, i}^{(3)}\rangle,\langle\tilde{T}_{3,i}^{(3)}\rangle,\cdots, \langle\tilde{T}_{N,i}^{(3)}\rangle)$ aligns with the highest weight $\omega_1$,
$U_i$s here are then equivalent to those obtained in the aGSWT.

\section{Continuous similarity transformation} 
\label{appCST}
To obtain the excitation spectra by the continuous similarity transformation~(CST)~\cite{PhysRevLett.115.207202},
we first express the Hamiltonian in terms of the bosonic operators $b_i, b_i^\dagger, c_i, c_i^\dagger$ via the transformation (\ref{ghp}). 
We keep up to quartic terms of bosonic creation and annihilation operators
After the Fourier transforms $b_i=\frac{1}{L}\sum_{\bm{k}}e^{i\bm{k} \cdot \bm{r}_i}b_{\bm{k}}$ and $c_i=\frac{1}{L}\sum_{\bm{k}}e^{i\bm{k} \cdot \bm{r}_i}c_{\bm{k}}$, and Bogoliubov transformations $b_{\bm{k}}=u^{b}_{\bm{k}}\beta_{\bm{k}}+v^{b}_{\bm{k}}\beta_{-\bm{k}}^{\dag}$ and $c_{\bm{k}}=u^{c}_{\bm{k}}\kappa_{\bm{k}}+v^{c}_{\bm{k}}\kappa_{-\bm{k}}^{\dag}$, the Hamiltonian takes the following form:
\begin{equation}
\begin{split}
	\mathcal{H} \approx \mathcal{H}_c+\mathcal{V}^{+}+\mathcal{V}^{-}.
\end{split}\label{HCST}
\end{equation}
Here, $\mathcal{H}_c$ encompasses all terms that conserve the boson number, while $\mathcal{V}^{+}$ and $\mathcal{V}^{-}$ are the hybridization terms that increase and decrease the particle number, respectively.
To be specific, $\mathcal{H}_c$, $\mathcal{V}^+$ and $\mathcal{V}^-$ are given by
\begin{widetext}
\begin{equation}
\begin{split}
	\mathcal{H}_c=&A+\sum_{1, 2}\{O_{1}(1, 2)\beta^{\dag}_{1}\beta_{2}+O_{2}(1, 2)\kappa^{\dag}_{1}\kappa_{2}\}\\
	+&\sum_{1234}\{C_1(1, 2, 3, 4)\beta_1^{\dag}\beta_2^{\dag}\beta_3\beta_4+C_3(1, 2, 3, 4)\kappa_1^{\dag}\kappa_2^{\dag}\kappa_3\kappa_4\}\\ 
	+&\sum_{1234}\{C_2(1, 2, 3, 4)\beta_1^{\dag}\beta_2\kappa_3^{\dag}\kappa_4+C_4(1, 2, 3, 4)\beta_1^{\dag}\beta_2^{\dag}\kappa_3\kappa_4+C_5(1, 2, 3, 4)\beta_1\beta_2\kappa_3^{\dag}\kappa_4^{\dag}\}.
\end{split}
\label{hnc}
\end{equation}
\begin{equation}
\begin{split}
\mathcal{V}^{+}=&\sum_{1, 2}\{O_{3}(1, 2)\beta^{\dag}_{1}\beta_{2}^{\dag}+O_{4}(1, 2)\kappa^{\dag}_{1}\kappa_{2}^{\dag}\}\\
	+&\sum_{1234}\{C_6(1, 2, 3, 4)\beta_1^{\dag}\beta_2^{\dag}\beta_3^{\dag}\beta_4+C_{9}(1, 2, 3, 4)\kappa_1^{\dag}\kappa_2^{\dag}\kappa_3^{\dag}\kappa_4
+C_{10}(1, 2, 3, 4)\beta_1^{\dag}\beta_2^{\dag}\beta_3^{\dag}\beta_4^{\dag}+C_{12}(1, 2, 3, 4)\kappa_1^{\dag}\kappa_2^{\dag}\kappa_3^{\dag}\kappa_4^{\dag}\}\\
	+&\sum_{1234}\{C_7(1, 2, 3, 4)\beta_1^{\dag}\beta_2^{\dag}\kappa_3^{\dag}\kappa_4+C_{8}(1, 2, 3, 4)\beta_1^{\dag}\beta_2\kappa_3^{\dag}\kappa_4^{\dag}
+C_{11}(1, 2, 3, 4)\beta_1^{\dag}\beta_2^{\dag}\kappa_3^{\dag}\kappa_4^{\dag}\}
\end{split}
\label{epc}
\end{equation}
\begin{equation}
\begin{split}
\mathcal{V}^{-}=&\sum_{1, 2}\{O_{5}(1, 2)\beta_{1}\beta_{2}+O_{6}(1, 2)\kappa_{1}\kappa_{2}\}\\
	+&\sum_{1234}\{C_{13}(1, 2, 3, 4)\beta_1^{\dag}\beta_2\beta_3\beta_4+C_{17}(1, 2, 3, 4)\beta_1\beta_2\beta_3\beta_4+C_{16}(1, 2, 3, 4)\kappa_1^{\dag}\kappa_2\kappa_3\kappa_4+C_{19}(1, 2, 3, 4)\kappa_1\kappa_2\kappa_3\kappa_4\}\\
	+&\sum_{1234}\{C_{14}(1, 2, 3, 4)\beta_1\beta_2\kappa_3^{\dag}\kappa_4+C_{15}(1, 2, 3, 4)\beta_1^{\dag}\beta_2\kappa_3\kappa_4+C_{18}(1, 2, 3, 4)\beta_1\beta_2\kappa_3\kappa_4\}.
\end{split}
\label{emc}
\end{equation}
\end{widetext}
Here, $A$ in Eq.~(\ref{hnc}) is a constant, and we simplify $\bm{k}_n$ to $n$, e.g., $\sum_{\bm{k}_1, \bm{k}_2}O_{1}(\bm{k}_1, \bm{k}_2)\beta^{\dag}_{\bm{k}_1}\beta_{\bm{k}_2}$ is rewritten as $\sum_{1, 2}O_{1}(1, 2)\beta^{\dag}_{1}\beta_{2}$. 
According to the CST theory, the Hamiltonian~(\ref{HCST}) depends on a continuous variable $l$.
The hybridization terms $\mathcal{V}^+$ and $\mathcal{V}^-$ are eliminated by solving the flow equation $\frac{\partial{\mathcal{H}(l)}}{\partial{l}}=[\mathcal{V}^+-\mathcal{V}^-, \mathcal{H}(l)]$,
ending up with $\mathcal{H}_{\textrm{eff}}$ when $l\rightarrow\infty$. Thus, the boson number is conserved in $\mathcal{H}_{\textrm{eff}}$ and it can be diagonalized in the subspace 
with a given particle number~\cite{10.21468/SciPostPhys.4.1.001}. For example, in the subspace with two bosons and momentum $\textbf{k}$, the basis can be divided into two sets. 
One is $\mathcal{B}_1$ comprising $|\beta_{\textbf{k}_1}^\dagger \beta_{\textbf{k}-\textbf{k}_1}^\dagger\rangle$, $|\kappa_{\textbf{k}_1}^\dagger \kappa_{\textbf{k}-\textbf{k}_1}^\dagger\rangle$ 
and the other $\mathcal{B}_2$ includes $|\beta_{\textbf{k}_1}^\dagger \kappa_{\textbf{k}-\textbf{k}_1}^\dagger\rangle$.
$\mathcal{H}_{\textrm{eff}}$ can be diagonalized under the basis $\mathcal{B}_1$ and $\mathcal{B}_2$, respectively. 
\bibliography{sun}
\end{document}